\documentclass[prd,twocolumn,preprintnumbers,showpacs,nofootinbib,notitlepage,floatfix]{revtex4-1}
\usepackage{hyperref}
\usepackage{graphicx}
\usepackage{amsmath}
\usepackage{amssymb}
\usepackage{slashed}
\usepackage{amsfonts}
\usepackage{xcolor}
\usepackage{multirow}
\usepackage[caption=false]{subfig}
\usepackage{array}
\usepackage{mwe}
\usepackage{mathrsfs}

\usepackage[utf8]{inputenc}
\usepackage[normalem]{ulem}

\DeclareRobustCommand{\eq}[1]{Eq.~\eqref{eq:#1}}

\DeclareRobustCommand{\app}[1]{App.~\ref{app:#1}}

\DeclareRobustCommand{\eq}[1]{Eq.~(\ref{eq:#1})}

\newcommand{\MS}{{\overline{\mathrm{MS}}}}

\newcommand{\eps}{\epsilon}

\newcommand{\nn}{\nonumber}

\newcommand\bets{\begin{table*}}
\newcommand\eets[1]{\label{tb:#1}\end{table*}}
\newcommand\scetii{SCET$_{\rm  II}$}

\begin{document}

\title{Transverse Momentum Distributions from Lattice QCD without Wilson Lines}

\author{Yong Zhao}
\email{yong.zhao@anl.gov}
\affiliation{Physics Division, Argonne National Laboratory, Lemont, IL 60439, USA}

\begin{abstract}
The transverse-momentum-dependent distributions (TMDs), which are defined by gauge-invariant 3D parton correlators with staple-shaped lightlike Wilson lines, can be calculated from quark and gluon correlators fixed in the Coulomb gauge on a Euclidean lattice. These quantities can be expressed gauge-invariantly as the correlators of Coulomb-gauge-dressed fields, which reduce to the standard TMD correlators under principal-value prescription in the infinite boost limit. In the framework of Large-Momentum Effective Theory, a quasi-TMD defined from such correlators in a large-momentum hadron state can be matched to the TMD via a factorization formula, whose exact form is derived using Soft Collinear Effective Theory and verified at one-loop order. Compared to the currently used gauge-invariant correlators, this new method can substantially improve statistical precision and simplify renormalization for the time-reversal-even TMDs, which will greatly enhance the predicative power of lattice QCD in the non-perturbative region.
\end{abstract}

\maketitle

The transverse-momentum-dependent parton distributions (TMDs) provide a 3D tomography of the nucleon in momentum space~\cite{Collins:1981uk,Collins:1981va,Collins:1984kg}, and are among the top targets for experiments at Fermilab, CERN, Jefferson Lab, RHIC and the forthcoming Electron-Ion Collider. Over the past two decades, significant progress has been made in the global fitting of quark TMDs from the semi-inclusive deep inelastic scattering and Drell-Yan processes. See the review in Ref.~\cite{Boussarie:2023izj}. The main goal there is to extract the intrinsic non-perturbative TMDs at parton transverse momentum $|k_\perp|\sim\Lambda_{\rm QCD}$, the most relevant region for nucleon structure, but the uncertainties in this domain still needs refined control.

Recently, the first-principles calculation in lattice quantum chromodynamics (QCD) has emerged as a powerful tool of making non-perturbative predictions for TMDs. 
Since TMDs are defined by quark and gluon correlators involving staple-shaped Wilson lines on the light-cone (LC), their explicit real-time dependence renders them impractical for direct lattice simulations.
Hence, initial efforts concentrated on the ratio of TMD moments, which are weighted averages in the longitudinal momentum fraction $x$ space and remain independent of time~\cite{Hagler:2009mb,Musch:2010ka,Musch:2011er,Engelhardt:2015xja,Yoon:2015ocs,Yoon:2017qzo}. Then, a breakthrough was made by Large-Momentum Effective Theory (LaMET)~\cite{Ji:2013dva,Ji:2014gla,Ji:2020ect} to calculate the $x$-dependence of parton distribution functions (PDFs), which has undergone profound development to make precision-controlled predictions nowadays~\cite{Gao:2021dbh,Zhang:2023bxs}. LaMET has also motivated the study of TMDs from lattice QCD~\cite{Ji:2014hxa,Ji:2018hvs,Ebert:2018gzl,Ebert:2019okf,Ebert:2019tvc,Ji:2019sxk,Ji:2019ewn,Vladimirov:2020ofp,Ebert:2020gxr,Ji:2020jeb,Ji:2021znw,Ebert:2022fmh,Rodini:2022wic,Schindler:2022eva,Deng:2022gzi,Zhu:2022bja,delRio:2023pse,Ji:2023pba}, leading to the calculations of the TMD evolution kernel~\cite{Shanahan:2020zxr,LatticeParton:2020uhz,Li:2021wvl,Schlemmer:2021aij,Shanahan:2021tst,LPC:2022ibr,Shu:2023cot,LatticePartonLPC:2023pdv,Avkhadiev:2023poz,Avkhadiev:2024mgd}, the soft function~\cite{LatticeParton:2020uhz,Li:2021wvl,LatticePartonLPC:2023pdv}, and their $(x,k_\perp)$ dependence~\cite{LatticePartonCollaborationLPC:2022myp,LatticeParton:2023xdl}.

In LaMET, one starts from a quasi-TMD, which is defined by a gauge-invariant equal-time correlator with a staple-shaped Wilson line in a large-momentum hadron state, and then perturbatively matches it to the TMD via a factorization formula~\cite{Ebert:2019okf,Ji:2019sxk,Ji:2019ewn,Ebert:2022fmh}.
Since this formalism was established, the lattice efforts have evolved from exploration to controlling the systematics. In lattice simulations, the staple-shaped Wilson line, constructed from gauge links with a substantial spatial extension, often imposes a notable memory overhead. More importantly, its self-energy contributes to an exponential suppression of the signal-to-noise ratio at large staple sizes, which correspond to the non-perturbative region essential for TMD physics. In a recent dedicated calculation of the evolution kernel~\cite{Avkhadiev:2024mgd}, the errors became too big at $|k_\perp|\lesssim (0.8\ {\rm fm})^{-1}$ to be meaningful. Besides, lattice renormalization must treat the linear power divergences from the Wilson lines~\cite{Ji:2017oey,Ishikawa:2017faj,Green:2017xeu} and the operator mixings due to discretization~\cite{Constantinou:2019vyb,Shanahan:2019zcq,Green:2020xco,Ji:2021uvr,Alexandrou:2023ucc}. Finally, the Wilson-line geometry incorporates multiple length scales, all of which must adhere to the hierarchy specified by the factorization formula to prevent intricate power corrections~\cite{Ebert:2019okf}.
Despite much progress in subtracting the linear divergences and operator mixings~\cite{Constantinou:2019vyb,Shanahan:2019zcq,Ebert:2019tvc,Green:2020xco,Ji:2021uvr,Zhang:2022xuw,Alexandrou:2023ucc,Avkhadiev:2023poz}, the remaining power corrections have not been fully analyzed.

In this work, we propose a method to calculate TMDs using quasi-TMDs defined by equal-time correlators fixed in the Coulomb gauge (CG), which is inspired by the authors' previous proposal for the PDFs~\cite{Gao:2023lny}.
The key distinguishing feature of our method is the absence of Wilson lines, which will save memory for their storage and, most importantly, substantially improve the statistical precision in lattice simulations. For the time-reversal-even (${\cal T}$-even) TMDs, it will allow access to the deep non-perturbative region that cannot reached by existing methods with the same resources. Furthermore, lattice renormalization becomes dramatically simplified without the linear divergences, operator mixings or power corrections resulting from the Wilson line and its imperfect geometry. Therefore, our proposal will greatly facilitate precise nucleon 3D imaging from lattice QCD.

This work is organized as follows. First, we introduce the CG quark quasi-TMD and re-express it gauge-invariantly as the correlation of CG-dressed quarks. Under an infinite Lorentz boost, the dressing factors reduce to lightlike Wilson lines, so the quasi-TMD correlator becomes the standard TMD correlator.
Next, we use Soft Collinear Effective Theory (SCET)~\cite{Bauer:2000ew,Bauer:2001ct,Bauer:2000yr,Bauer:2001yt} to derive the exact factorization formula that relates the quasi-TMD to the TMD, which also leads to the definition of the quasi soft function. We verify the factorization at one-loop order in perturbation theory and calculate the matching coefficient. Finally, we show how to determine the quasi soft function through a meson form factor~\cite{Ji:2019sxk} and the CG quasi-TMD wave function, thus completing the recipe for TMD calculation on the lattice.

\vspace{0.3cm} 
\paragraph*{Definition.} 
The CG quasi-TMD is defined as
\begin{align}\label{eq:qtmd}
    \tilde B(x,b_\perp,\mu,P^z) &\equiv P^z\int_{-\infty}^\infty \frac{db^z}{2\pi} e^{i x P^z b^z} \\
    &\quad \times {1\over 2P^t}\langle P| \bar{\psi}(b)\gamma^t \psi(0)\Big|_{\vec{\nabla}\cdot \vec{A}=0}|P\rangle\,,\nn
\end{align}
where $b^\mu=(0,\vec{b})=(0,b_\perp,b^z)$, $|P\rangle$ is a hadron state with momentum $P^\mu=(P^t,0,0,P^z)$ normalized to $\langle P|P\rangle = 2P^t\delta^{(3)}(0)$, and $\mu$ is the $\MS$ scale.
The CG condition $\vec{\nabla}\cdot \vec{A}=0$ is fixed so that the correlator can have a nonzero matrix element, and its renormalization only involves the quark wave function which is multiplicative and free from linear power divergence~\cite{Zwanziger:1998ez,Baulieu:1998kx,Andrasi:2005xu,Niegawa:2006ey,Niegawa:2006hg,Gao:2023lny}. Note that the CG condition does not completely fix the gauge due to the residual gauge degrees of freedom and Gribov copies~\cite{Gribov:1977wm,Singer:1978dk}. On the lattice, the former issue manifests as the CG being independently fixed on each time slice, so it does not pose a problem for equal-time correlators. As for the Gribov copies, numerical studies show that they contribute to a fluctuation which is typically smaller than the gauge noise~\cite{Giusti:2001xf,Kalusche:2024osk,Gao:2024fbh}, so they are not a concern within practical lattice precisions. Finally, due to 3D rotational symmetry, $\vec{P}$ can be oriented to any direction, which allows for access to larger off-axis momenta.

Under an infinite Lorentz boost along the $z$-direction, the CG approaches the LC gauge $A^+=0$~\cite{Ji:2013fga,Hatta:2013gta}, so it is easy to see the connection between the quasi- and LC TMDs. Nevertheless, it is still illuminating to show in a gauge-invariant manner how the staple-shaped lightlike Wilson lines emerge under the infinite boost, which also turns out to be useful for the SCET derivation of factorization formula below.
To start with, we look for a gauge-invariant dressing of the quark field~\cite{Lavelle:1995ty}, which was first studied by Dirac for U(1) gauge theory~\cite{Dirac:1955uv},
\begin{align}\label{eq:dress}
	\Psi_C(b) &= U_C(b) \psi(b)\,.
\end{align}
The factor $U_C(b)$ is a gauge transformation that rotates an unfixed gauge potential $\vec{A}$ to the CG~\cite{Lavelle:1995ty,Zhao:2015kca},
\begin{align}\label{eq:Ac}
	\vec{A}_C &\equiv U_C \vec{A} U_C^{-1} + {i\over g} U_C \vec{\nabla} U_C^{-1}\,,\quad \vec{\nabla}\cdot \vec{A}_C=0\,,
\end{align}
where $g$ is the strong coupling.
Under an arbitrary gauge transformation $U$, the dressing factor $U_C$ transforms as
\begin{align}
	U_C \longrightarrow U_C U^{-1}
\end{align}
if we require $U(b)$ to be trivial at spatial boundaries, which do not contribute to parton physics at finite $x$ after an infinite boost.
Therefore, $\Psi_C(b)$ is gauge invariant, and so is the quasi-TMD correlator given by
\begin{align}\label{eq:cgi}
O^C_{\gamma^t}(b)&=\bar{\Psi}_C(b)\gamma^t \Psi_C(0)\,.
\end{align}

The solution to $U_C$ is not unique due to the Gribov copies~\cite{Lavelle:1995ty}. Nevertheless, they are infrared (IR) effects and should be considered in lattice gauge fixing. For deriving the factorization relation between quasi- and LC TMDs, we can focus on the perturbative solution,
\begin{align}\label{eq:omega}
	U_C &= 1+\sum_{m=1}^\infty {(ig)^m\over m!} \pmb{\omega}_m\,.
\end{align}
where $\pmb{\omega}=\omega_a t^a$ and $t^a$ is an SU(3) generator. 
Using the unitarity relation, $U_C^{-1}=U_C^\dagger$, we obtain the unique compact solution to \eq{Ac} as
\begin{align}
	 \pmb{\omega}_1 &=-{1\over \nabla^2} \vec{\nabla}\cdot \vec{A}\,,\qquad\qquad\qquad \ldots \,,\nn\\
	 {\pmb{\omega}_{m+1}\over (m+1)!}&= {1\over \nabla^2 }\sum_{k=0}^{m}{1\over k!}{(-1)^{k}\over (m-k)!}\left\{{1\over k+1}\vec{\nabla}\cdot \left( \pmb{\omega}^\dagger_{k+1} \vec{\nabla} \pmb{\omega}_{m-k}\right) \right.\nn\\
	 &\left.\quad  - \left[\pmb{\omega}^\dagger_k \vec{\nabla} \pmb{\omega}_{m-k}, \cdot \vec{A} \right] \right\}\,,\qquad \ldots\,.
\end{align}
Notably, if we fix $\vec{\nabla}\cdot \vec{A}=0$, then $\pmb{\omega}_{m}=0$, and $U_C=\mathbf{1}$. See more details of the solution $U_C$ in \app{uc}.

Next, we boost $\pmb{\omega}_m$. Since our goal is to identify the leading-power contribution in the effective theory expansion of the quasi-TMD, the Lorentz boost is performed at classical level, which is in the same spirit as deriving the Heavy Quark Effective Theory and SCET Lagrangians.
In the infinite boost limit, $|\partial^+| \gg |\partial^\perp| \gg |\partial^-|$, so
\begin{align}\label{eq:exp}
	\pmb{\omega}_1(b)&\rightarrow  {\partial^+ \over (\partial^+)^2 - 0} A^+ (b)   \equiv {1\over \partial^+_{\rm{pv}}}A^+(b)\\
	&=-{1\over 2} \left[ \int_{b^-}^{-\infty^-} + \int_{b^-}^{+\infty^-}  \right]db_1^-\ A^+(b^+,b_1^-,  b_\perp)\,,\nn
\end{align}
where the LC coordinates $b^\pm=(b^t\pm b^z)/\sqrt{2}$, and the principal-value (PV) prescription of LC singularity, $1/\partial^+_{\rm{pv}}$, naturally arises from the CG gluon propagator. Then, we can prove iteratively that
\begin{align}\label{eq:omegan}
    {\pmb{\omega}_m\over m!}   &\rightarrow {1\over \partial^+_{\rm{pv}}}\Big(\! \cdots\! \Big({1\over \partial^+_{\rm{pv}}} \Big( \big({1\over \partial^+_{\rm{pv}}} \overbrace{A^+\big) A^+\Big)A^+\Big)\!\cdots\! A^+}^{m}\Big)\,.
\end{align}
If $1/\partial^+_{\rm{pv}}$ is replaced with the advanced or retarded prescriptions, then $\pmb{\omega}_m$ becomes a path-ordered integral from $b^-$ to $\mp \infty^-$, so $U_C$ reduces to a lightlike Wilson line,
\begin{align}\label{eq:colimit}
	U_C &\!\rightarrow\! {\cal W}^\dagger_{\pm n}(b) \!\equiv\! \mathscr{P}\exp\!\left[-ig\! \int^{\mp\infty^-}_{b^-}\! db_1^- A^+(b_1^-)\right]\,.
\end{align}
More details of the infinite boost transformation are provided in \app{imf}. We are unsure about the form of a PV-regularized Wilson line, so we denote it as ${\cal W}_n^\dagger$ for bookkeeping purposes. Thus, $O^C_{\gamma^t}(b)$ becomes the standard TMD correlator,
\begin{align}
    O^C_{\gamma^t}(b)&\rightarrow {O_{\gamma^+}(b)\over\sqrt{2}}={1\over\sqrt{2}}\bar{\psi}(b){\cal W}_{n}(b) \gamma^+ {\cal W}^\dagger_{n}(0)\psi(0) \,,
\end{align}
with the Dirac structure $\gamma^t = (\gamma^+ + \gamma^-)/\sqrt{2}$ dominated by the plus component.

The ${\cal T}$-even TMDs~\cite{Mulders:1995dh}, including the unpolarized, helicity, transversity, wormgear and pretzelosity distributions, are independent of the Wilson-line direction or LC regularization, so they can all be accessed by the CG quasi-TMDs. However, this universality is broken for the ${\cal T}$-odd Sivers and Boer-Mulders functions~\cite{Sivers:1989cc,Boer:1997nt,Collins:2002kn,Brodsky:2002cx,Brodsky:2002rv}. To access them we must introduce an additional dressing factor or condition that breaks the parity between $+z$ and $-z$ directions.
If we fix the CG, which corresponds to fixing TMDs in the LC gauge, then one might think of inserting a transverse gauge link at $b^z=\pm \infty$ to define the ${\cal T}$-odd functions~\cite{Ji:2002aa,Belitsky:2002sm}. However, such an insertion is exponentially suppressed as the CG is nonsingular in the spatial directions. Therefore, the longitudinal Wilson lines seem inevitable, but that is exactly the gauge-invariant approach~\cite{Ji:2020jeb}. See also \app{trans}. Nevertheless, since the transverse link at LC infinity is a zero-mode effect~\cite{Ji:2020baz}, there may still be hope in exploiting the topological properties of gauge configurations to avoid Wilson lines, which will preserve the computational advantages of our method. Finding the solution is beyond the scope of this work and will be studied in the future.

\vspace{0.3cm}

\paragraph*{Factorization.} For a hadron moving at a large momentum $P$ along the $z$-direction, its wave function is dominated by collinear quark and gluon modes whose momentum $k^\mu$ scales as $(k^+,k^-,k_\perp)=(1,\lambda^2,\lambda)P^+$ with $\lambda \ll 1$. When the transverse momentum $k_\perp^\mu$ is probed by a TMD, the soft modes $(\lambda,\lambda,\lambda)P^+$ become relevant, and the system can be naturally described by type-II Soft-Collinear Effective Theory (SCET$_{\rm  II}$)~\cite{Bauer:2002uv,Bauer:2001yt,Beneke:2002ph,Bauer:2003mga}, which enables us to derive the factorization formula connecting the CG quasi-TMD in QCD to the TMD in SCET. Besides, the gauge-invariant formulation of quasi-TMD correlators allows us to work in the covariant gauge.

In \scetii, the relevant degrees of freedom in QCD fields are collinear ($n$), soft ($s$) and ultra soft ($us$),
\begin{align}
	\psi &= \psi_n + \psi_s + \psi_{us} + \ldots\,,\nn\\
	 A^\mu &= A_n^\mu + A_s^\mu + A_{us}^\mu + \ldots\,,
\end{align}
where $\ldots$ are hard modes to be integrated over and softer modes that are further suppressed. The quark and gluon modes scale as $\{\psi_n,\psi_s,\psi_{us}\}\sim \{\lambda,\lambda^{3/2},\lambda^3\}(P^+)^{3/2}$, $A_n^\mu= (A^+_n, A^-_n, A^\perp_n) \sim (1,\lambda^2,\lambda)P^+$, and $\{ A^\mu_s ,A^\mu_{us}\}\sim \{\lambda, \lambda^2\}P^+$~\cite{Bauer:2001yt,Bauer:2002uv}.
Since $\bar{\psi}_n\gamma^- \psi_n \sim O(\lambda^4)$, $\bar{\psi}_n\gamma^+ \psi_n \sim O(\lambda^2)$, the leading-power expansion of $O^C_{\gamma^t}(b)$ is
\begin{align}\label{eq:colexp}
	O^C_{\gamma^t}(b) = {1\over \sqrt{2}}\bar{\psi}_n(b) U^\dagger_C \gamma^+ U_C \psi_n(0) + O(\lambda^4)\,.
\end{align}
The dressing factor $U_C$ includes all gluon modes as they contribute at $O(1)$. But thanks to the decoupling of collinear and soft modes in the SCET Lagrangian~\cite{Bauer:2001yt}, it can be decomposed in a path integral as
\begin{align}
	U_C(A) &= U_C(A_s)U_C(A_n) \equiv U^s_CU^n_C
\end{align}
after we integrate out the $O(\sqrt{\lambda}P^+)$ virtual fluctuations. Since only the collinear and soft modes are considered, we did not further separate the ultrasoft modes from them, which is why $U_C(A_{us})$ is absent here.

According to \eq{colimit}, $U^n_C$ can be expanded as
\begin{align}\label{eq:Uexp}
	U^n_C&= \mathscr{P} \exp\left[{g\over {\cal P}^+} A_n^+\right] + O(\lambda^n) \equiv W^\dagger_n + O(\lambda^n)\,,
\end{align}
where ${\cal P}^\mu$ is the Hermitian derivative operator that picks out the collinear (label) momentum, and $W^\dagger_n(A_n)$ is the collinear Wilson line in SCET. 
At classical level, the power corrections $O(\lambda^n)$ start at $n=2$, but due to the energy-dependence of $U_C$, there could be $n=1$ corrections at higher loop orders~\cite{Liu:2023onm}.
In contrast, $U^s_C$ cannot be expanded like \eq{exp} or \eq{Uexp} because the soft momentum scales as $(\partial^x, \partial^y, \partial^z)\sim (\lambda,\lambda,\lambda)P^+$.

Therefore,  in the collinear expansion $U_C\psi_n$ is matched to the leading gauge-invariant SCET operator as~\cite{Bauer:2001yt,Bauer:2002nz}
\begin{align}
	U_C\psi_n(b) &= e^{-i{\cal P}\cdot b}C({\cal P}^+/\mu)\left[U^s_C S_n W_n^\dagger\xi_n\right](b)\,,
\end{align}
where $\xi_n$ is the collinear quark field. The soft Wilson line $S_n = \mathscr{P} \exp\big[-{g\over {\cal P}^-} A_s^-\big]$
must be added to ensure SCET gauge symmetry~\cite{Bauer:2001yt}. The matching coefficient $C({\cal P}^+/\mu)$ is flavor independent and real under PV prescription.

Since $b^z$ and $b_\perp$ are Fourier conjugate to $xP^z=xP^+/\sqrt{2}$ and $k_\perp$, respectively, we have the scaling $b^z\sim O(1)$ and $b_\perp\sim O(1/\lambda)$ at finite $x$. Therefore, the exchange of hard particles between collinear fields, separated by $b_\perp$, is suppressed, and the CG quasi-TMD correlator can be matched to SCET as
\begin{align}\label{eq:fact0}
	&{e^{i{\cal P}\cdot b}\over\sqrt{2}}\Big\langle P\Big| \left[\bar{\xi}_nW_n S^\dagger_n (U^{s}_C)^\dagger\right]\!(b)\ C({\cal P}^+/\mu)^\dagger\ \gamma^+\nn\\
	&\quad \times C({\cal P}^+/\mu)\left[ U^s_C S_n W_n^\dagger\xi_n\right]\!(0) \Big|P \Big\rangle\,.
\end{align}
Moreover, we can have the multipole expansion in $b^z$,
\begin{align}
	U^{s}_C(b) &= U^{s}_C(b_\perp) + b^z \partial_z U^{s}_C(b_\perp) + \ldots \approx U^{s}_C(b_\perp) \,,
\end{align}
for $\partial_z U^{s}_C\sim O(\lambda)$. Since the collinear and soft modes decouple, \eq{fact0} can be factorized as
\begin{align}
	&{e^{i{\cal P}\cdot b}\over\sqrt{2}} \Big\langle P\Big|\! \left[\bar{\xi}_nW_n \right]\!(b) C\big({{\cal P}^+\over\mu}\big)^\dagger\gamma^+ C\big({{\cal P}^+\over\mu}\big)\!\left[ W_n^\dagger\xi_n\right]\!(0)\Big|P \Big\rangle \nn\\
	&\quad \times{1\over N_c} \langle 0| \mbox{Tr}\left[ S^\dagger_n(b_\perp) U^{s\dagger}_C(b_\perp) U^s_C(0) S_n(0)\right] |0 \rangle\,,
\end{align}
where $N_c=3$. Here the hadronic matrix element is the beam function $B$ with zero-bin subtracted~\cite{Manohar:2006nz,Stewart:2009yx}, and the vacuum matrix element defines a soft function
\begin{align}
S_C^0(b_\perp) \!\equiv\! {1\over N_c} \langle 0| \mbox{Tr}\!\left[ S^\dagger_n(b_\perp) U^{s\dagger}_C(b_\perp) U^s_C(0) S_n(0)\right]\! |0 \rangle
\end{align}
which can be understood as the zero-bin of $\tilde B$. After Fourier transform, we obtain the factorization formula
\begin{align}
	\tilde B(x,b_\perp,\mu,P^z) &= |C(xP^+/\mu)|^2 B(x,b_\perp,\ldots, xP^+) \nn\\
	&\qquad \times S_C^0(b_\perp,\ldots)\,,
\end{align}
where $\ldots$ stands for the ultraviolet (UV) and rapidity regulators~\cite{Ebert:2019okf}. The factor ${1/\sqrt{2}}$ is absorbed into the integration measure. Both $B$ and $S_C^0$ contain rapidity divergences that cancel between each other, and an overall UV renormalization is implied here and below.

The TMD $f$ is obtained from $B$ with the combination of a soft function $S$~\cite{Ebert:2019okf},
\begin{align}
	f(x, b_\perp,\mu,\zeta) &= {B(x,b_\perp,\ldots, xP^+) }S(b_\perp,\ldots,y_n)\,,
\end{align}
where the rapidity divergences cancel, leaving a dependence on the Collins-Soper scale $\zeta=2(xP^+)^2e^{-2y_n}$~\cite{Collins:1981uk,Collins:1981va} that evolves with the parameter $y_n$ introduced by $S$.
Hence, the factorization formula for ${\cal T}$-even TMDs is
\begin{align}\label{eq:fact1}
	{\tilde B(x, b_\perp, \mu, P^z) \over \tilde S_C(b_\perp,\mu,y_n)} &= |C(xP^+/\mu)|^2 f(x, b_\perp, \mu, \zeta)\,,
\end{align}
where the quasi soft function $\tilde S_C \equiv S_{C}^0/S$ is free from rapidity divergence and satisfies the evolution equation
\begin{align}
	{d\over dy_n}\ln \tilde S_C(b_\perp,\mu,y_n) &= K(b_\perp,\mu) \,,
\end{align}
where $K(b_\perp,\mu)$ is the Collins-Soper kernel~\cite{Collins:1981uk,Collins:1981va}. Therefore, we can go one-step further by setting $y_n=0$, and the factorization formula becomes
\begin{align}\label{eq:fact2}
	&{\tilde B(x, b_\perp, \mu, P^z) \over \tilde S_C(b_\perp,\mu,0)} = |C(xP^+/\mu)|^2 \\
 &\qquad \times \exp\left[{1\over 2}K(b_\perp,\mu) \ln{2(xP^+)^2\over \zeta}\right]f(x, b_\perp, \mu, \zeta)\,,\nn
\end{align}
the same as gauge-invariant quasi-TMDs~\cite{Ebert:2019okf,Ji:2019sxk,Ji:2019ewn,Ebert:2022fmh}. More intermediate steps of the SCET derivation can be found in \app{fact}.

To verify the above factorization, we perform a perturbative calculation for a free massless quark target at momentum $p^\mu=(p^z,0,0,p^z)$, with UV and IR divergences regulated by dimensional regularization ($d=4-2\eps$).
See \app{oneloop} for the explicit diagrammatic evaluations. At one-loop order, the $\MS$ quasi-TMD and soft function are
\begin{align}
	&\tilde B^{(1)}(x, b_\perp,\mu, p^z,\eps) \\
    &= {\alpha_sC_F\over 2\pi} \left[-{1+x^2\over 1-x }\left({1\over\eps}+ L_b\right) + (1-x) \right]_+^{(0,1)} \nn\\
    & + \delta(1-x){\alpha_sC_F\over 2\pi}\left[ {5\over2}L_b - 3L_p - {(L_b\!-\!L_p)^2\over 2}  - {23\over2} + {\pi^2\over2}\right]\,,\nn\\
	&\tilde S_C^{(1)}(b_\perp, \mu, y_n) = {\alpha_sC_F\over 2\pi} (1-2y_n) L_b \,,
\end{align}
where $\alpha_s=g^2/(4\pi)$, $C_F=4/3$, $L_b = \ln(\mu^2 b_\perp^2 e^{2\gamma_E}/4)$, $L_p = \ln{\mu^2/(4p_z^2)}$, and $[\ldots]_+^{(0,1)}$ denotes a plus function within $(0,1)$.
The subtracted TMD is~\cite{Ebert:2019okf}
\begin{align}
	&f^{(1)}(x, b_\perp,\mu,\zeta,\eps) \\
    &=  {\alpha_sC_F\over 2\pi} \left[-{1+x^2\over 1-x }\left({1\over\eps}+ L_b\right) + (1-x) \right]^{(0,1)}_+ \nn\\
    &+ \delta(1-x){\alpha_sC_F\over 2\pi}\left[ - {L_b^2\over2} + L_b\left({3\over2} + \ln{\mu^2\over \zeta}\right) + {1\over2} - {\pi^2\over12}\right]\,,\nn
\end{align}
where $\zeta=4p_z^2 e^{-2y_n}$. Therefore, the difference
\begin{align}
	&\tilde B^{(1)} -\delta(1-x) \tilde S_C^{(1)} - f^{(1)} \nn\\
 &= \delta(1-x) {\alpha_sC_F\over 2\pi}\left[-{L^2_p\over 2}-3L_p-12+{7\pi^2\over12}\right]
\end{align}
is free from the IR logarithm $L_b$, which validates the factorization~\cite{Ebert:2019okf}.
As a result, we also obtain
\begin{align}
    C\Big({p^z\over\mu}\Big) &\!=\! 1 \!+\! {\alpha_s C_F\over 4\pi}\left[-{L^2_p\over 2}-3L_p-12+{7\pi^2\over12}\right] \,.
\end{align}

\vspace{0.3cm}
\paragraph*{Soft function.} Though the factorization formula for the CG quasi-TMD has been established, the soft function $\tilde S_C$, which involves lightlike Wilson lines, is still not directly calculable on the lattice. For the gauge-invariant quasi-TMD, one method was proposed to extract the soft function from a light-meson form factor~\cite{Ji:2019sxk}
\begin{align}
    F(b_\perp,P^z) &= \langle \pi(-\vec{P}) | j_1(b_\perp) j_2(0) | \pi(\vec{P})\rangle\,, 
\end{align}
where $\pi(\vec{P})$ can be a pion with momentum $\vec{P}=(0,0,P^z)$, and $j_{1,2}=\bar{\psi}\Gamma_{1,2}\psi$ are light-quark currents.

At large $P^z$, $F$ satisfies a Drell-Yan-like factorization formula~\cite{Ji:2019sxk,Ji:2021znw,Deng:2022gzi},
\begin{align}\label{eq:fact_ff}
    F(b_\perp,P^z) &= \int dx_1 dx_2\ H_F(x_1, x_2, P^z,\mu)\\
    & \times \phi^\dagger (x_1,b_\perp, \mu, P^+,y_n) \phi(x_2,b_\perp,\mu, P^+,-y_n)\,, \nn
\end{align}
where $\phi$ is a TMD wave function, and $y_n$-dependence cancels between the $\phi$'s. The coefficient $H_F$ is available at one-loop~\cite{Ji:2019sxk,Deng:2022gzi}.

We can construct a CG quasi-TMD wave function $\tilde \phi$ which is matched to a PV-regularized wave function as
\begin{align}\label{eq:fact_wf}
	{\tilde \phi(x, b_\perp, \mu,\! P^z) \over \tilde S(b_\perp,\mu,y_n)} &\!=\! C\big({xP^+\over\mu}\!\big)C\big({\bar{x}P^+\over \mu}\!\big) \phi_{\rm pv}(x, b_\perp, \mu, \zeta)\,,
\end{align}
where $\bar{x}=1-x$. Without loss of generality, we can set $y_n=0$ and plug the above result into \eq{fact_ff}, then
\begin{align}
    F(b_\perp,P^z) &= \int dx_1 dx_2{H_F(x_1, x_2, P^z,\mu)\over C(x_1) C(x_2) C(\bar{x}_1) C(\bar{x}_2)} \\
    &\qquad \times {\tilde \phi(x_1, b_\perp, \mu, P^z) \over \tilde S_C(b_\perp,\mu,0)} {\tilde \phi(x_2, b_\perp, \mu, P^z) \over \tilde S_C(b_\perp,\mu,0)}\,.\nn
\end{align}
which can be used to extract $\tilde S_C$.
Since the CG wave function is real~\cite{Belitsky:2002sm}, there is no need to show the cancellation of imaginary parts~\cite{LPC:2022ibr,LatticePartonLPC:2023pdv,Avkhadiev:2023poz}, which provides another useful simplification for lattice analysis. 
This completes our new method for TMD calculation.

\vspace{0.3cm}
\paragraph*{Conclusion.} We have proposed a new method for calculating the TMDs from CG correlators. Through a gauge-invariant extension, we demonstrate that they reduce to the standard TMD correlators under PV prescription in the infinite boost limit. Moreover, using SCET we derived the factorization formula for the quasi-TMD and the definition of the quasi soft function. The latter can be extracted from a meson form factor and the CG quasi-TMD wave function, which completes our framework for calculating the TMDs. With the absence of Wilson lines, we expect this method to significantly improve the precision of ${\cal T}$-even TMDs and play a major role in nucleon 3D imaging from lattice QCD.

Going forward, we plan to pursue a more rigorous proof of factorization with the leading region method~\cite{Collins:1989gx} and the solution for the ${\cal T}$-odd functions without Wilson lines. It is straightforward to extend our method to Wigner distributions by assigning proper parton momenta to each matching coefficient $C$, like in \eq{fact_wf}. The gluon sector follows similar derivations, except that we should address operator mixings in the CG and the Gribov copies, which will be studied in future works.

\begin{acknowledgments}
We thank Xiang Gao, Rui Zhang, Yushan Su, Jinchen He, Xiangdong Ji, Iain Stewart, Stella Schindler, Michael Wagman, Anthony Grebe, Phiala Shanahan, Artur Avkhadiev, Swagato Mukherjee, George Flemming and Yizhuang Liu for valuable discussions.
This material is based upon work supported by the U.S. Department of Energy, Office of Science, Office of Nuclear Physics through Contract No.~DE-AC02-06CH11357, the \textit{Quark-Gluon Tomography (QGT) Topical Collaboration} with Award DE-SC0023646, and the 2023 Physical Sciences and Engineering (PSE) Early Investigator Named Award program at Argonne National Laboratory.

\end{acknowledgments}

\appendix

\section{Solution to $U_C$}
\label{app:uc}

Firstly, we note that the dressing factor $U_C$ is only introduced in the continuum theory for studying the boost limit of the Coulomb-gauge correlator, which also allows us to formally derive the factorization formula using SCET. On the Euclidean lattice, we fix the correlators in the Coulomb gauge, which makes the numerical simulation much more precise.

For an unfixed gauge potential $\vec{A}$, the dressing factor $U_C(\vec{A})$, which is also a gauge transformation, satisfies the condition
\begin{align}\label{eq:coulomb}
	\vec{\nabla} \cdot \left[ U_C \vec{A} U_C^{-1} + {i\over g} U_C \vec{\nabla} U_C^{-1}\right] &=0\,.
\end{align}

Under an arbitrary gauge transformation,
\begin{align}
    \vec{A} &\longrightarrow U\vec{A} U^{-1} + {i\over g} U\vec{\nabla}U^{-1}\,,
\end{align}
we have
\begin{align}
    &\vec{\nabla} \cdot \left[ U'_C \vec{A}' U'^{-1}_C + {i\over g} U'_C \vec{\nabla} U'^{-1}_C\right]\nn\\
    &\!=\!\vec{\nabla}\! \cdot\! \left[U'_C \Big(U\vec{A} U^{-1} \!+\! {i\over g} U\vec{\nabla}U^{-1}\Big)U'^{-1}_C\right] \!+\! {i\over g}\vec{\nabla}\cdot\left[ U'_C \vec{\nabla} U'^{-1}_C \right]\nn\\
    &\!=\! \vec{\nabla} \cdot \left[U'_C U\vec{A} (U'_C U)^{-1}\right] \nn\\
    &\qquad + {i\over g} \vec{\nabla} \cdot \left[U'_CU\vec{\nabla}U^{-1} U'^{-1}_C + U'_C \vec{\nabla} U'^{-1}_C \right]\nn\\
    &\!=\! \vec{\nabla} \cdot \left[U'_C U\vec{A} (U'_C U)^{-1}\right] + {i\over g} \vec{\nabla} 
    \cdot \left[ (U'_CU) \vec{\nabla} (U'_CU)^{-1}\right] \nn\\
    &=0.
\end{align}
For arbitrary $\vec{A}$, the solution to the above equation would require $U_C$ to transform as
\begin{align}
    U'_C &= U_C U^{-1}\,, 
\end{align}
so that the above equation is automatically satisfied according to \eq{coulomb}. As a result, the combination $U_C\psi$ is gauge invariant.

An exception could be when both $U_C$ and $U$ are independent of the spatial coordinates, i.e., $\partial^i U_C=\partial^i U=0$, which could happen at the spatial infinities if we allow the gauge potential to be non-vanishing there. In this case, there can still be residual gauge degrees of freedom in $U_C\psi$. However, under an infinite boost the spatial infinity will approach the light-cone infinity, which will not affect the PDF or TMD at finite $x$. Besides, though on the lattice the Coulomb gauge is fixed independently on each time slice, it does not pose a problem for the equal-time correlators that we consider.

Now let us find the solution to \eq{coulomb}, which can be rewritten as
\begin{align}\label{eq:coulomb2}
	 -{i\over g} \vec{\nabla}\cdot (U_C^{-1}\vec{\nabla} U_C)+ [ U_C^{-1}\vec{\nabla} U_C, \cdot \vec{A}] + \vec{\nabla}\cdot \vec{A} &=0\,.
\end{align}

Due to the existence of Gribov copies in the Coulomb gauge, the solution to $U_C$ is not unique. However, such effects are infrared and should be considered in lattice gauge fixing. For the derivation of factorization formula, we can focus on the perturbative solution to $U_C$, which is unique for compact gauge configurations.

Let us do a perturbative expansion of $U_C$ as
\begin{align}
	U_C &= \sum_{n=0}^\infty {(ig)^n\over n!} \pmb{\omega}_n\,,
\end{align}
where $\pmb{\omega}_0=1$.
Because $U_C$ is unitary, we also have
\begin{align}
	U_C^{-1} &= U_C^\dagger = \sum_{n=0}^\infty {(-ig)^n\over n!} \pmb{\omega}^\dagger_n\,.
\end{align}
Therefore,
\begin{align}
	U_C^{-1}\vec{\nabla} U_C &= \sum_{m=0}^\infty\sum_{n=0}^\infty {(-ig)^m\over m!}{(ig)^n\over n!} \pmb{\omega}^\dagger_m \vec{\nabla} \pmb{\omega}_n \nn\\
    &= \sum_{n=0}^\infty \sum_{m=0}^n {(ig)^n(-1)^m\over m!(n-m)!}\pmb{\omega}^\dagger_m \vec{\nabla} \pmb{\omega}_{n-m} \nn\\
	&= \sum_{n=0}^\infty (ig)^n\sum_{m=0}^{n} {(-1)^m\over m!(n-m)!}\pmb{\omega}^\dagger_m \vec{\nabla} \pmb{\omega}_{n-m}\,.
\end{align}
Plugging the above into \eq{coulomb2}, we have order by order in $g$
\begin{align}
	\nabla^2 \pmb{\omega}_1 &=-\vec{\nabla}\cdot \vec{A}\,,\nn\\
	{1\over 2!} \nabla^2 \pmb{\omega}_2 &=\vec{\nabla}\cdot \left(\pmb{\omega}^\dagger_1 \vec{\nabla}\pmb{\omega}_1 \right) - [ \vec{\nabla} \pmb{\omega}_1, \cdot \vec{A}]\,,\nn\\
	 {1\over 3!}\nabla^2\pmb{\omega}_3  &={1\over 2} \vec{\nabla}\cdot  \left( - \pmb{\omega}^\dagger_2 \vec{\nabla}\pmb{\omega}_1 + \pmb{\omega}^\dagger_1 \vec{\nabla}\pmb{\omega}_2   \right) \nn\\
     &\qquad - \big[ - \pmb{\omega}^\dagger_1 \vec{\nabla}\pmb{\omega}_1 + {1\over 2}\vec{\nabla}\pmb{\omega}_2, \cdot \vec{A} \big]\,,\nn\\
	 &\ldots \nn\\
	 {1\over (n+1)!}\nabla^2 \pmb{\omega}_{n+1}&= \sum_{m=0}^{n}{1\over m!}{(-1)^{m}\over (n-m)!}\nn\\
     &\times \left\{{1\over m+1}\vec{\nabla}\cdot \left( \pmb{\omega}^\dagger_{m+1} \vec{\nabla} \pmb{\omega}_{n-m}\right) \right.\nn\\
     &\qquad \qquad \left.- \left[\pmb{\omega}^\dagger_m \vec{\nabla} \pmb{\omega}_{n-m}, \cdot \vec{A} \right] \right\}\,,\nn\\
        &\ldots
\end{align}

For compact gauge fields and transformations, we have the unique solutions
\begin{align}
	 \pmb{\omega}_1 &=-{1\over \nabla^2} \vec{\nabla}\cdot \vec{A}\,,\nn\\
	{\pmb{\omega}_2\over 2!}   &={1\over \nabla^2}\left( \vec{\nabla}\cdot \left(\pmb{\omega}^\dagger_1 \vec{\nabla}\pmb{\omega}_1 \right) - [ \vec{\nabla} \pmb{\omega}_1, \cdot \vec{A}]\right)\,,\nn\\
	 {\pmb{\omega}_3\over 3!}  &={1\over \nabla^2} \left({1\over 2} \vec{\nabla}\cdot  \left( - \pmb{\omega}^\dagger_2 \vec{\nabla}\pmb{\omega}_1 + \pmb{\omega}^\dagger_1 \vec{\nabla}\pmb{\omega}_2   \right) \right.\nn\\
     &\qquad \left.- \big[ - \pmb{\omega}^\dagger_1 \vec{\nabla}\pmb{\omega}_1 + {1\over 2}\vec{\nabla}\pmb{\omega}_2, \cdot \vec{A} \big]\right)\,,\nn\\
	 &\ldots \nn\\
	 {\pmb{\omega}_{n+1}\over (n+1)!} &={1\over \nabla^2} \sum_{m=0}^{n}{1\over m!}{(-1)^{m}\over (n-m)!}\nn\\
     &\times \left\{{1\over m+1}\vec{\nabla}\cdot \left( \pmb{\omega}^\dagger_{m+1} \vec{\nabla} \pmb{\omega}_{n-m}\right)\right.\nn\\
     &\qquad \qquad \left.- \left[\pmb{\omega}^\dagger_m \vec{\nabla} \pmb{\omega}_{n-m}, \cdot \vec{A} \right] \right\}\,,\nn\\
    &\ldots
\end{align}

\section{Infinite boost limit}
\label{app:imf}

In this section, we demonstrate that $U_C$ approaches the light-cone Wilson line under an infinite Lorentz boost along the $z$-direction. Note that the purpose of studying this limit is to identify the leading-power operator in the effective theory expansion of the quasi-TMD correlator, so the Lorentz transformation is performed at classical level. This is in the same spirit as deriving the leading-power HQET or SCET Lagrangian in the infinite heavy quark mass or collinear limit.

At classical level the operator can always be expressed interchangeably between coordinate and momentum spaces through a Fourier transform,
\begin{align}
    {1\over \nabla^2} \vec{\nabla}\cdot \vec{A}(x) &= -i\int{d^4 k\over (2\pi)^4}\ e^{-ik\cdot x} {1\over \vec{k}^2}\vec{k}\cdot  \vec{\tilde A}(k)\,.
\end{align}
If we boost the operator to a frame where $k^0\sim k^z \gg k_\perp$ and $A^0\sim A^z \gg A_\perp$, or equivalently, $k^+ \gg k_\perp \gg k^-$ and $A^+ \gg A_\perp \gg A^-$, then 
\begin{align}\label{eq:imf}
    \pmb{\omega}_1&=-{1\over \nabla^2} \vec{\nabla}\cdot \vec{A}(x) \nn\\
    &=i\int{d^4 k\over (2\pi)^4}\ e^{-ik\cdot x} {1\over k_z^2+k_\perp^2}\left[k^z \tilde A^z(k)+\vec{k}_\perp\cdot \vec{\tilde A}_\perp(k)\right]\nn\\
    &\approx  i\int{d^4 k\over (2\pi)^4}\ e^{-ik\cdot x} {k^z\over k_z^2 + k_\perp^2} \tilde A^z(k) \nn\\
    & \approx  i\int{d^4 k\over (2\pi)^4}\ e^{-ik\cdot x} {k^+\over (k^+)^2+\eps^2}\tilde A^+(k)\nn\\
    &={1\over 2} \left[ \int^{x^-}_{-\infty^-} + \int^{x^-}_{+\infty^-}  \right]d\eta^-\ A^+(x^+,\eta^-,  x_\perp)\nn\\
    &\equiv {1\over \partial^+_{\rm{pv}}}A^+(x)\,,
\end{align}
which is exactly the principal-valued prescription of the inverse partial derivative $1/\partial^+$ and corresponds to an anti-periodic boundary condition at light-cone infinities $\pm \infty^-$~\cite{Ji:2002aa,Belitsky:2002sm},
\begin{align}\label{eq:antip}
    \pmb{\omega}_1 (\infty^-) & = - \pmb{\omega}_1 (-\infty^-)\,.
\end{align}

Plugging \eq{imf} into $\pmb{\omega}_2$, we have
\begin{align}
    {1\over 2!}  \pmb{\omega}_2 &={1\over (\partial^+_{\rm{pv}})^2}\left[ \partial^+ \left(\pmb{\omega}^\dagger_1 \partial^+\pmb{\omega}_1 \right) - [\partial^+ \pmb{\omega}_1, A^+]\right] \nn\\
    & = {1\over \partial^+_{\rm{pv}}} \left[ \big({1\over \partial^+_{\rm{pv}}}A^+\big) A^+\right]\,,
\end{align}
where we have used $\partial^+\pmb{\omega}_1 = A^+$ and $(A^+)^\dagger = A^+$.

With the above result, we can prove iteratively that
\begin{align}\label{eq:omegan}
    {1\over n!}  \pmb{\omega}_n &={1\over \partial^+_{\rm{pv}}}\left( \ldots \left({1\over \partial^+_{\rm{pv}}} \left( \big({1\over \partial^+_{\rm{pv}}}A^+\big) A^+\right)A^+\right)\ldots A^+\right)
\end{align}
by showing that the infinite boost limit of $\pmb{\omega}_{n+1}$,
\begin{align}
	& {\pmb{\omega}_{n+1}\over (n+1)!} ={1\over (\partial^+_{\rm{pv}})^2} \sum_{m=0}^{n}{1\over m!}{(-1)^{m}\over (n-m)!}\\
     &\quad \times \left\{{1\over m+1}\partial^+ \left( \pmb{\omega}^\dagger_{m+1} \partial^+ \pmb{\omega}_{n-m}\right) \!-\! \left[\pmb{\omega}^\dagger_m \partial^+ \pmb{\omega}_{n-m}, A^+ \right] \right\}\,,\nn
\end{align}
is equivalent to the form in \eq{omegan}. 
 
 Since
\begin{align}
	{\partial^+\pmb{\omega}_n\over n!} &= {\pmb{\omega}_{n-1}\over (n-1)!} A^+\,,\nn\\
    {\partial^+\pmb{\omega}^\dagger_n\over n!} &=(A^+)^\dagger {\pmb{\omega}^\dagger_{n-1}\over (n-1)!}=A^+{\pmb{\omega}^\dagger_{n-1}\over (n-1)!} \,,
\end{align}
we have
\begin{align}
	&\sum_{m=0}^{n}{1\over m!}{(-1)^{m}\over (n-m)!}\left[\pmb{\omega}^\dagger_m \partial^+ \pmb{\omega}_{n-m}, A^+ \right] \nn\\
	&=\sum_{m=0}^{n}{1\over m!}{(-1)^{m}\over (n-m)!}\left[\pmb{\omega}^\dagger_m \partial^+ \pmb{\omega}_{n-m} A^+ - A^+\pmb{\omega}^\dagger_m \partial^+ \pmb{\omega}_{n-m}\right]\nn\\
	&=\sum_{m=0}^{n}{1\over m!}{(-1)^{m}\over (n-m)!}\left[ -\partial^+\pmb{\omega}^\dagger_m  \pmb{\omega}_{n-m} A^+ - A^+\pmb{\omega}^\dagger_m \partial^+ \pmb{\omega}_{n-m}\right]\nn\\
	&=\sum_{m=0}^{n}{1\over m!}{(-1)^{m}\over (n-m)!}\left[-{\partial^+\pmb{\omega}^\dagger_m \partial^+ \pmb{\omega}_{n-m+1}\over n-m+1} \right.\nn\\
	&\qquad \qquad \left. - {\partial^+\pmb{\omega}^\dagger_{m+1} \partial^+ \pmb{\omega}_{n-m}\over m+1}\right]\nn\\
	&=- \sum_{m=0}^{n}{1\over m!}{(-1)^{m}\over (n-m+1)!}{\partial^+\pmb{\omega}^\dagger_m \partial^+ \pmb{\omega}_{n-m+1}} \nn\\
    &\qquad - \sum_{m=0}^{n}{1\over (m+1)!}{(-1)^{m}\over (n-m)!}{\partial^+\pmb{\omega}^\dagger_{m+1} \partial^+ \pmb{\omega}_{n-m}}\nn\\
	&=- \sum_{m=0}^{n}{1\over m!}{(-1)^{m}\over (n-m+1)!}{\partial^+\pmb{\omega}^\dagger_m \partial^+ \pmb{\omega}_{n-m+1}} \nn\\
    &\qquad- \sum_{m=1}^{n+1}{1\over m!}{(-1)^{m-1}\over (n-m+1)!}{\partial^+\pmb{\omega}^\dagger_{m} \partial^+ \pmb{\omega}_{n-m+1}}\nn\\
    &=- \sum_{m=1}^{n}{1\over m!}{(-1)^{m}\over (n-m+1)!}{\partial^+\pmb{\omega}^\dagger_m \partial^+ \pmb{\omega}_{n-m+1}} \nn\\
    &\qquad+ \sum_{m=1}^{n}{1\over m!}{(-1)^{m}\over (n-m+1)!}{\partial^+\pmb{\omega}^\dagger_{m} \partial^+ \pmb{\omega}_{n-m+1}}\nn\\  
	&=0\,,
\end{align}
where we have used $\partial^+\pmb{\omega}^\dagger_0 = \partial^+\pmb{\omega}_0=0$. To derive the third line we have used the unitarity relation $U_C^\dagger U_C=1$ to obtain the identities
\begin{align}\label{eq:unit}
	\sum_{m=0}^{n} {(-1)^m\over m!(n-m)!}\pmb{\omega}^\dagger_m \pmb{\omega}_{n-m} &= 0 \,,\\
    \sum_{m=0}^{n} {(-1)^m \pmb{\omega}^\dagger_m \partial^+\pmb{\omega}_{n-m}\over m!(n-m)!} &= - \sum_{m=0}^{n} {(-1)^m \partial^+\pmb{\omega}^\dagger_m \pmb{\omega}_{n-m}\over m!(n-m)!}\,.
\end{align}
In addition, we can also prove iteratively that the unitarity relation in \eq{unit} is satisfied by the solution in \eq{omegan}, i.e.,
\begin{align}
    &\partial^+_{\rm pv}\sum_{m=0}^{n+1} {(-1)^m\over m!(n+1-m)!}\pmb{\omega}^\dagger_m \pmb{\omega}_{n+1-m}\nn\\ 
    &= \sum_{m=1}^{n+1} {(-1)^m\over (m-1)!(n+1-m)!} A^+\pmb{\omega}^\dagger_{m-1} \pmb{\omega}_{n+1-m} \nn\\
    &\qquad + \sum_{m=0}^{n} {(-1)^m\over m!(n-m)!}\pmb{\omega}^\dagger_m \pmb{\omega}_{n-m}A^+\nn\\
    &= -A^+ \sum_{m=0}^{n} {(-1)^m\over m!(n-m)!} \pmb{\omega}^\dagger_{m} \pmb{\omega}_{n-m} + 0 \nn\\
    &=0\,.
\end{align}
Using the principal-value prescription for the inverse derivative, we have
\begin{align}
    \sum_{m=0}^{n+1} {(-1)^m\over m!(n+1-m)!}\pmb{\omega}^\dagger_m \pmb{\omega}_{n+1-m} &={1\over \partial_{\rm pv}^+}0 = 0\,.
\end{align}

After all, as $\pmb{\omega}_n$ is nothing but the $n$-th order term in the perturbative expansion of a path-ordered Wilson line, the unitariy relation in \eq{unit} must hold.

Finally, by combining the above results, we have
\begin{align}
    &{\pmb{\omega}_{n+1}\over (n+1)!} \nn\\
    &={1\over (\partial^+_{\rm{pv}})^2} \sum_{m=0}^{n}{1\over m!}{(-1)^{m}\over (n-m)!}{1\over m+1}\partial^+ \left( \pmb{\omega}^\dagger_{m+1} \partial^+ \pmb{\omega}_{n-m}\right)\nn\\
    &={1\over \partial^+_{\rm{pv}}} \sum_{m=1}^{n+1}{1\over m!}{(-1)^{m-1}\over (n-m+1)!} \left( \pmb{\omega}^\dagger_{m} \partial^+ \pmb{\omega}_{n+1-m}\right)\nn\\
    &={1\over \partial^+_{\rm{pv}}} \sum_{m=1}^{n}{1\over m!}{(-1)^{m-1}\over (n-m)!}\left( \pmb{\omega}^\dagger_{m} \pmb{\omega}_{n-m}A^+\right)    \nn\\
    &=-{1\over \partial^+_{\rm{pv}}}\sum_{m=0}^{n}{1\over m!}{(-1)^{m-1}\over (n-m)!}\left( \pmb{\omega}^\dagger_{m} \pmb{\omega}_{n-m}A^+\right) + {1\over \partial^+_{\rm{pv}}}\left(\pmb{\omega}_nA^+\right)\nn\\
    &={1\over \partial^+_{\rm{pv}}}\left(\pmb{\omega}_nA^+\right)\,.
\end{align}

This completes our proof that \eq{omegan} is the solution to $\pmb{\omega}_n$ in the infinite boost limit.

Now let us take a closer look at this solution. If we, instead of choosing the principal-value prescription, define the inverse derivative as
\begin{align}
    \left[{1\over \partial^+ }\right]_\pm A^+(x) &\equiv \int^{x^-}_{\mp\infty^-} dy^- A^+(y^-)\,,
\end{align}
then
\begin{align}
	 {1\over n!}  \pmb{\omega}_n^{\mp} &= \int_{\mp \infty^-}^{x^-} dy_1^- \int_{\mp \infty^-}^{y^-_1} dy_2^-\cdots \int_{\mp \infty^-}^{y^-_{n-1}} dy_n^- \nn\\
	 &\qquad \times  A^+(y_n^-)\cdots A^+(y_2^-)A^+(y_1^-)\nn\\
	 & = (-1)^n  \int^{\mp \infty^-}_{x^-} dy_1^- \int^{\mp \infty^-}_{y^-_1} dy_2^-\cdots \int^{\mp \infty^-}_{y^-_{n-1}} dy_n^- \nn\\
	 &\qquad \times A^+(y_n^-)\cdots A^+(y_2^-)A^+(y_1^-)\,,
\end{align}
which is exactly the path-ordered integration from $x^-$ to $\mp\infty^-$.

As a result, the dressing factor
\begin{align}\label{eq:imf2}
	U_C^{\mp} &= \sum_{n=0}^\infty  {(ig)^n\over n!}  \pmb{\omega}_n^{\mp}\nn\\
	&= \sum_{n=0}^\infty  (-ig)^n \int^{\mp \infty^-}_{x^-} dy_1^- \int^{\mp \infty^-}_{y^-_1} dy_2^-\cdots \int^{\mp \infty^-}_{y^-_{n-1}} dy_n^- \nn\\
	&\qquad \times A^+(y_n^-)\cdots A^+(y_2^-)A^+(y_1^-)\nn\\
	&= {\cal P} \exp\left[ -ig \int^{\mp \infty^-}_{x^-} dy^- A^+(y^-)\right] \equiv W_{\pm n}^\dagger
\end{align}
becomes a lightlike Wilson line pointing to $\mp \infty^-$.

When the inverse derivative is defined with principal-value prescription, $U_C$ can be regarded as a certain ``average'' of past- and future-pointing Wilson lines. Although we are not sure if it can be expressed in the form of standard Wilson lines, this quantity is well defined as all the ${\cal T}$-even TMDs correspond to the principal-value prescription. Therefore, we will denote it as $W_n^\dagger$ for simplicity.

\section{Origin of transverse link}
\label{app:trans}

As we have shown above, the infinite boost limit of the Coulomb gauge quasi-TMD is the light-cone TMD under principal-value prescription. In the light-cone gauge $A^+=0$, in order to define the ${\cal T}$-odd TMDs under principal-value prescription, one has to insert a transverse link at past or future light-cone infinity~\cite{Belitsky:2002sm}. Therefore, naively one would also propose to introduce a transverse gauge link at $\pm \infty n_z$ to define the ${\cal T}$-odd quasi-TMDs\,,
\begin{align}\label{eq:t-odd2}
	O^C_{\gamma^t}(b)\Big|_{\vec{\nabla}\cdot \vec{A}=0}&= \bar{\psi}(b){\cal W}_{\perp}(\pm\infty n^z; b_\perp, 0_\perp) \gamma^t \psi(0)\,.
\end{align}

However, the above conjecture has oversimplified the relation between spatial and light-cone infinities. Under an infinite Lorentz boost along the $z$-direction, all finite $z$ are boosted to the light-cone infinity, so it is not clear that the original spatial infinity $z=\pm\infty$ has any direct relation to the latter. Instead, since the Coulomb-gauge is not singular in the 3D spatial directions, it might well be that matrix elements involving such a transverse link is exponentially suppressed in distance.

To verify this, we have calculated the one-loop correction to a quasi-TMD wave function in a pion with quark and antiquark carrying momentum fraction $x_0$ and $\bar{x}_0=1-x_0$, respectively. It is well known that the gauge-invariant TMD wave function has an imaginary part whose sign depends on the orientation of the Wilson lines. If we fix the light-cone gauge $A^+=0$, then the transverse link at $\pm \infty^-$ will contribute to the imaginary part, or more accurately a phase~\cite{Belitsky:2002sm}. Therefore, a simple exercise for us to is check whether the transverse link at $\pm \infty n^z$ in the Coulomb gauge will have the same imaginary part.

According to our one-loop result, the imaginary part of the Coulomb gauge quasi-TMD wave function defined above is zero. We have also calculated it with the transverse link at a finite distance $L$ away, and found that the imaginary part is suppressed as $L$ approaches $\pm \infty$. Since in this work we focus only on the ${\cal T}$-even TMDs, we leave out the details of this one-loop calculation here.

Considering that the Coulomb gauge is not singular in the 3D spatial directions, it is probably still necessary to introduce the longitudinal Wilson lines that break the parity between $+z$ and $-z$ directions. However, this will lead to the exactly same gauge-invariant staple-shaped Wilson line operator that already exists in the literature, so there is no computational advantage here. Nevertheless, we still have hope that there may exist a novel operator in the Coulomb gauge without Wilson lines, which will give rise to the ${\cal T}$-odd TMDs, thus maintaining the computational advantages of our new method. This, however, cannot be accomplished within the time frame of this work, but will be actively pursued in the future.

\section{SCET derivation of factorization}
\label{app:fact}

For a hadron moving at a large momentum $P$ along the $z$-direction, its wave function is dominated by collinear quark and gluon modes whose momentum $k^\mu$ scales as $(k^+,k^-,k_\perp)=(1,\lambda^2,\lambda)P^+$ with $\lambda \ll 1$. When the transverse momentum $k_\perp^\mu$ is probed by a TMD, the soft modes $(\lambda,\lambda,\lambda)P^+$ become relevant, and the system can be naturally described by type-II Soft-Collinear Effective Theory (SCET$_{\rm  II}$)~\cite{Bauer:2002uv,Bauer:2001yt,Beneke:2002ph,Bauer:2003mga}, which allows us to derive the factorization formula connecting the CG quasi-TMD in QCD to the TMD in SCET.

Our strategy is to first identify the leading SCET operator under the collinear expansion of the Coulomb-gauge correlator, and then use mode separation and the matching between QCD and SCET operators to derive the exact form of the factorization formula. Note that the collinear expansion is done at classical level, and the nontrivial dynamics is encoded in the matching coefficients and power corrections.
To take advantage of the spacetime and gauge symmetries of \scetii, we express the Coulomb-gauge correlator gauge-invariantly with the dressing factor $U_C$, so that we can work in the covariant gauge.

In \scetii, the relevant degrees of freedom are the collinear ($n$), soft ($s$) and ultra soft ($us$) ones. The QCD field can be decomposed as
\begin{align}
	\psi &= \psi_n + \psi_s + \psi_{us} + \ldots\,,\nn\\
	 A^\mu &= A_n^\mu + A_s^\mu + A_{us}^\mu + \ldots\,,\nn
\end{align}
where $\ldots$ stands for the hard modes to be integrated over and softer modes that are further suppressed by powers of $\lambda$. The quark and gluon modes scale as $\{\psi_n,\psi_s,\psi_{us}\}\sim \{\lambda,\lambda^{3/2},\lambda^3\}(P^+)^{3/2}$, $A_n^\mu= (A^+_n, A^-_n, A^\perp_n) \sim (1,\lambda^2,\lambda)P^+$, and $\{ A^\mu_s ,A^\mu_{us}\}\sim \{\lambda, \lambda^2\}P^+$~\cite{Bauer:2001yt,Bauer:2002uv}.
Since $\bar{\psi}_n\gamma^- \psi_n \sim O(\lambda^4)$, $\bar{\psi}_n\gamma^+ \psi_n \sim O(\lambda^2)$, and $\gamma^t=(\gamma^++\gamma^-)/\sqrt{2}$, the quark bilinear $O^C_{\gamma^t}(b)=\bar{\psi}(b)U_C^\dagger \gamma^t U_C\psi(0)$ can be expanded as
\begin{align}\label{eq:colexp}
	O^C_{\gamma^t}(b) = {1\over \sqrt{2}}\bar{\psi}_n(b) U^\dagger_C \gamma^+ U_C \psi_n(0) + O(\lambda^4)\,.
\end{align}
The dressing factor $U_C$ includes all the gluon modes as they contribute at $O(1)$. But thanks to the decoupling of collinear and soft modes in the SCET Lagrangian~\cite{Bauer:2001yt}, when evaluated in the path integral, it can be decomposed as
\begin{align}
	U_C(A) &= U_C(A_s)U_C(A_n) \equiv U^s_CU^n_C
\end{align}
after integrating out the $O(\sqrt{\lambda}P^+)$ virtual fluctuations. Note that since only the collinear and soft modes are of our concern, we did not further separate the ultrasoft modes from them, which is why $U_C(A_{us})$ is absent here. If we perform a field redefinition of the collinear modes in SCET~\cite{Bauer:2002nz}, then $U_C(A_{us})$ will emerge.

According to the infinite boost limit of $U_C$ in \eq{imf}, $U^n_C$ can be expanded at tree-level as the collinear Wilson line up to power corrections,
\begin{align}
	U^n_C(A_n)&=W^\dagger_n + O(\lambda^2)\,.
\end{align}
In contrast, $U^s_C$ cannot be expanded in this way because 
\begin{align}
    {1\over \nabla^2} \vec{\nabla}\cdot \vec{A}_s(x) &= -i\int{d^4 k_s\over (2\pi)^4}\ e^{-ik_s\cdot x} {1\over \vec{k}_s^2}\vec{k}_s\cdot  \vec{\tilde A}_s(k_s)\,.
\end{align}
scales homogeneously in the momentum space, i.e., $\vec{k}_s=(k^x,k^y,k^z)\sim (\lambda, \lambda,\lambda)P^+$. Therefore, we must keep $U^s_C$ in the collinear expansion.

As a result, in the collinear limit,
\begin{align}
    U_C\psi_n &=U^s_C W^\dagger_n \psi_n \,.   
\end{align}
In SCET, the r.h.s. corresponds to
\begin{align}
       U^s_C  \left[e^{-i{\cal P}\cdot b} W^\dagger_n \xi_n \right] \,,
\end{align}
where ${\cal P}^\mu$ is the Hermitian derivative operator that picks out the collinear (label) momentum of $W^\dagger_n \xi_n$, and $\xi_n$ is the collinear quark field with the zero mode subtracted~\cite{Manohar:2006nz}. However, the above operator is not gauge invariant in SCET.

Under a collinear gauge transformation $\mathcal{U}$,
\begin{align}
    \xi_n \rightarrow \mathcal{U} \xi_n\,, \quad W_n^\dagger \rightarrow W_n^\dagger \mathcal{U}^\dagger\,,\quad A_s^\mu \rightarrow A_s^\mu\,,
\end{align}
so the operator $U^s_C  W^\dagger_n \xi_n $ is gauge invariant.

Under a soft gauge transformation $V_s$,
\begin{align}
    \xi_n \rightarrow \xi_n\,, \quad W_n^\dagger \rightarrow W_n^\dagger \,,\quad A_s^\mu \rightarrow V_s \left(A_s^\mu + {1\over g}\cal{P}\right)V_s^{-1}\,,
\end{align}
so the dressing factor $U_C^s$ transforms as $U_C^s\to U_C^sV_s^{-1}$ according to its definition, and 
\begin{align}
    U^s_C  W^\dagger_n \xi_n \rightarrow U^s_C V_s^{-1} W^\dagger_n \xi_n
\end{align}
is not gauge invariant. Therefore, we must insert a soft Wilson line~\cite{Bauer:2001yt}
\begin{align}
    S_n = \mathscr{P} \exp\big[-{g\over {\cal P}^-} A_s^-\big]
\end{align}
which emerges after integrating out $O(\sqrt{\lambda}P^+)$ off-shell fluctuations and transforms as
\begin{align}
    S_n &\rightarrow V_s S_n\,,
\end{align}
so that the new operator 
\begin{align}
    U^s_C S_n W^\dagger_n \xi_n \rightarrow U^s_C S_n W^\dagger_n \xi_n
\end{align}
is gauge invariant.

Therefore, $U_C\psi_n$ is matched to the gauge-invariant SCET operator as~\cite{Bauer:2001yt,Bauer:2002nz}
\begin{align}
	U_C\psi_n(b) &= e^{-i{\cal P}\cdot b}C({\cal P}^+/\mu)\left[U^s_C S_n W_n^\dagger\xi_n\right](b)\,,
\end{align}
where $C({\cal P}^+/\mu)$ is a matching coefficient and independent of the quark flavor, and $e^{-i{\cal P}\cdot b}$ can be moved to the front of the operator because it only acts on the collinear fields.

Since $b^z$ and $b_\perp$ are Fourier conjugate to $xP^z=xP^+/\sqrt{2}$ and $k_\perp$, respectively, we have the scaling $b^z\sim O(1)$ and $b_\perp\sim O(1/\lambda)$ at finite $x$. Therefore, the exchange of hard particles between collinear fields, which are separated by $b_\perp$, is suppressed, and the Coulomb-gauge quasi-TMD correlator can be multiplicatively matched to SCET through the matching of the dressed fields,
\begin{align}\label{eq:fact0}
	& {1\over \sqrt{2}}e^{i{\cal P}\cdot b}\Big\langle P\Big| \left[\bar{\xi}_nW_n S^\dagger_n (U^{s}_C)^\dagger\right]\!(b)\ C({\cal P}^+/\mu)^\dagger \gamma^+\nn\\
	&\qquad \qquad  \times C({\cal P}^+/\mu)\left[ U^s_C S_n W_n^\dagger\xi_n\right]\!(0) \Big|P \Big\rangle\,.
\end{align}
Moreover, we can have the multipole expansion in $b^z$ since $\partial^z U_C^s \sim O(\lambda)$,
\begin{align}
	U^{s}_C(b) &= U^{s}_C(b_\perp) + b^z \partial_z U^{s}_C(b_\perp) + \ldots = U^{s}_C(b_\perp) + O(\lambda)\,.
\end{align}
Using the separation of collinear and soft modes in SCET, we can factorize \eq{fact0} as
\begin{align}
	&\sum_{i,j}^{N_c}{e^{i{\cal P}\cdot b}\over\sqrt{2}} \Big\langle \!P\Big|\! \left[\bar{\xi}_nW_n \right]_i\!(b) C(\!{{\cal P}^+\over\mu}\!)^\dagger \gamma^+  C(\!{{\cal P}^+\over\mu}\!)\!\left[ W_n^\dagger\xi_n\right]_j\!(0)\Big|P\! \Big\rangle \nn\\
	&\qquad\times \langle 0| \left[ S^\dagger_n(b_\perp) (U^{s}_C)^\dagger(b_\perp) U^s_C(0) S_n(0)\right]_{ij} |0 \rangle\nn\\
 &=\!\sum_{i,j}^{N_c}\!{e^{i{\cal P}\cdot b}\over\sqrt{2}}\! \Big\langle\! P\Big|\! \left[\bar{\xi}_nW_n\! \right]_i\!(b) C(\!{{\cal P}^+\over\mu}\!)^\dagger \gamma^+ C(\!{{\cal P}^+\over\mu}\!)\!\left[\! W_n^\dagger\xi_n\!\right]_j\!(0)\Big|P \!\Big\rangle \nn\\
	&\qquad\times{1\over N_c} \langle 0| \mbox{Tr}\left[ S^\dagger_n(b_\perp) (U^{s}_C)^\dagger(b_\perp) U^s_C(0) S_n(0)\right] |0 \rangle \delta_{ij}\nn\\
 &={e^{i{\cal P}\cdot b}\over\sqrt{2}} \Big\langle \!P\Big|\! \left[\bar{\xi}_nW_n \right]\!(b) C\big({{\cal P}^+\over\mu}\big)^\dagger\gamma^+ C\big({{\cal P}^+\over\mu}\big)\!\left[ W_n^\dagger\xi_n\right]\!(0)\Big|P\! \Big\rangle \nn\\
	&\quad\times{1\over N_c} \langle 0| \mbox{Tr}\left[ S^\dagger_n(b_\perp) (U^{s}_C)^\dagger(b_\perp) U^s_C(0) S_n(0)\right] |0 \rangle 
\end{align}
where $N_c=3$, and we used the unitarity of SU(3) gauge transformation to obtain the soft function as the trace of a color matrix. Here the hadronic matrix element is the beam function $B$ with the zero-bin subtracted~\cite{Manohar:2006nz,Stewart:2009yx}, and the vacuum matrix element defines a soft function
\begin{align}
S_C^0(b_\perp) \equiv {1\over N_c} \langle 0| \mbox{Tr}\left[ S^\dagger_n(b_\perp) (U^{s}_C)^\dagger(b_\perp) U^s_C(0) S_n(0)\right] |0 \rangle\,,
\end{align}
which can be understood as the zero-bin of $\tilde B$. It is worth mentioning that if we fix the Coulomb gauge, $S_C^0(b_\perp)$ becomes the vacuum expectation value of a pair of open-ended collinear Wilson lines.

After Fourier transform, we obtain the factorization formula in $x$-space,
\begin{align}
	&\tilde B(x,b_\perp,\mu,P^z)\nn\\
	&\equiv \int {db^z\over 2\pi}e^{ixP^z b^z} {1\over\sqrt{2}}e^{i{\cal P}\cdot b} \Big\langle P\Big|\! \left[\bar{\xi}_nW_n \right]\!(b) C\big({{\cal P}^+/\mu}\big)^\dagger\gamma^+\nn\\
	&\qquad\times C\big({{\cal P}^+/\mu}\big)\!\left[ W_n^\dagger\xi_n\right]\!(0)\Big|P \Big\rangle\ S_C^0(b_\perp)\nn\\
    &=\Big\langle P\Big|\! \left[\bar{\xi}_nW_n \right]\! C\big({{\cal P}^+/\mu}\big)^\dagger\gamma^+ {1\over\sqrt{2}}\delta ({\cal P}^z- xP^z) \nn\\
    &\qquad \times C\big({{\cal P}^+/\mu}\big)\!\left[ W_n^\dagger\xi_n\right]\!\Big|P \Big\rangle\ S_C^0(b_\perp)\nn\\
    &=\Big\langle P\Big|\! \left[\bar{\xi}_nW_n \right]\! C\big({{\cal P}^+/\mu}\big)^\dagger\gamma^+ \delta ({\cal P}^+- xP^+) \nn\\
    &\qquad \times C\big({{\cal P}^+/\mu}\big)\!\left[ W_n^\dagger\xi_n\right]\!\Big|P \Big\rangle S_C^0(b_\perp)\nn\\ 
    &= |C(xP^+/\mu)|^2 B(x,b_\perp,\ldots, xP^+)  S_C^0(b_\perp,\ldots)\,,
\end{align}
where $\ldots$ stands for the ultraviolet (UV) and rapidity regulators~\cite{Ebert:2019okf}. The factor ${1/\sqrt{2}}$ in \eq{colexp} is absorbed by the integration measure. Both $B$ and $S_C^0$ contain rapidity divergences that cancel between each other, and an overall UV renormalization is implied here and below.

The TMD $f$ is obtained from zero-bin subtracted $B$ with the combination of a soft function $S$~\cite{Ebert:2019okf},
\begin{align}
	f(x, b_\perp,\mu,\zeta) &= {B(x,b_\perp,\ldots, xP^+) }S(b_\perp,\ldots,y_n)\,,
\end{align}
where the rapidity divergences cancel with a remnant dependence on the Collins-Soper scale $\zeta=2(xP^+)^2e^{-2y_n}$ that evolves with the parameter $y_n$ introduced in $S$~\cite{Collins:1981uk,Collins:1981va}.

Hence, the full factorization formula is
\begin{align}\label{eq:fact1}
	{\tilde B(x, b_\perp, \mu, P^z) \over \tilde S_C(b_\perp,\mu,y_n)} &= |C(xP^+/\mu)|^2 f(x, b_\perp, \mu, \zeta)\,,
\end{align}
where the quasi soft factor $\tilde S_C \equiv S_{C}^0/S$ is free from rapidity divergence. The quasi soft factor satisfies the Collins-Soper evolution equation~\cite{Collins:1981uk,Collins:1981va},
\begin{align}
	{d\over dy_n}\ln \tilde S_C(b_\perp,\mu,y_n) &= K(b_\perp,\mu) \,,
\end{align}
where $K(b_\perp,\mu)$ is the Collins-Soper kernel. Therefore, we can go one-step further by setting $y_n=0$ in $ \tilde S_C$ and re-express the factorization formula in \eq{fact1} as
\begin{align}
	&{\tilde B(x, b_\perp, \mu, P^z) \over \tilde S_C(b_\perp,\mu,0)} = |C(xP^+/\mu)|^2\\
	&\qquad \times \exp\left[{1\over 2}K(b_\perp,\mu) \ln{2(xP^+)^2\over \zeta}\right]f(x, b_\perp, \mu, \zeta)\,,\nn
\end{align}
which is in similar form as the gauge-invariant quasi-TMDs~\cite{Ebert:2019okf,Ji:2019sxk,Ji:2019ewn,Ebert:2022fmh}.

\section{One-loop calculation}
\label{app:oneloop}

\subsection{Quasi beam function}
The quasi-TMD in the Coulomb gauge is defined as
\begin{align}
	\tilde q(x, b_\perp, P^z) &={P^z\over P^t} \int {db^z \over 4\pi} e^{ix P^z b^z}\langle P | \bar{\psi}(b) \gamma^t  \psi(0)\Bigg|_{\vec{\nabla}\cdot \vec{A}=0} |P\rangle\,.
\end{align}

The gluon propagator in this gauge is
\begin{align}
	iD^{\mu\nu}(k) & = {-i\over k^2+i0} \left[g^{\mu\nu} -  n\cdot k{n^\mu k^\nu + n^\nu k^\mu \over \vec{k}^2} + {k^\mu k^\nu \over \vec{k}^2}\right] \nn\\
	&\equiv  {-i\over k^2+i0} \pi^{\mu\nu}(k)\,,
\end{align}
where $n^\mu=(1,0,0,0)$.

The real part contribution to the one-loop quasi-TMD in an on-shell massless quark state with momentum $p^\mu=(p^z,0,0,p^z)$ is
\begin{align}
\tilde{q}_{\rm real}(b,p) &= \bar{u}(p) \int {d^dk\over (2\pi)^d} (-ig\gamma_\mu\tau^a) {i\over \slashed k } \gamma^0 {i\over\slashed k} (-ig\gamma_\nu\tau^a)\nn\\
&\qquad \times iD^{\mu\nu}(p-k) u(p) e^{ik\cdot b}\nn\\
&= -ig^2 C_F \int {d^dk\over (2\pi)^d} \frac{\mbox{Tr}\left[\gamma_\mu\slashed k \gamma^0 \slashed k \gamma_\nu \slashed p \right]/2}{k^4 (p-k)^2} \nn\\
&\qquad \times \pi^{\mu\nu}(p-k) e^{ik\cdot b}\,.
\end{align}
We use Ward identity to define the on-shell wave function renormalization, i.e., the virtual contribution, so the full one-loop correction to the quasi-TMD is
\begin{align}
\tilde{q}(b,p) 
&= -ig^2 C_F \int {d^dk\over (2\pi)^d} \frac{\mbox{Tr}\left[\gamma_\mu\slashed k \gamma^0 \slashed k \gamma_\nu \slashed p \right]/4}{k^4 (p-k)^2}\nn\\
&\qquad \times \pi^{\mu\nu}(p-k) \left(e^{ik\cdot b} - e^{ip\cdot b}\right)\,.
\end{align}

The Feynman gauge part is
\begin{align}
&\tilde{q}_1(b,p) \\
&= -ig^2 C_F \int {d^dk\over (2\pi)^d} \frac{\mbox{Tr}\left[\gamma_\mu\slashed k \gamma^0 \slashed k \gamma^\mu \slashed p \right]/4}{k^4 (p-k)^2} \left(e^{ik\cdot b} - e^{ip\cdot b}\right)\nn\\
&= -i(D\!-\!2)g^2 C_F\! \int {d^dk\over (2\pi)^d} \left[\frac{p^0 \!-\! k^0}{k^2 (p\!-\!k)^2} \!-\! \frac{k^0}{k^4} \right] \left(e^{ik\cdot b} \!-\! e^{ip\cdot b}\right)\nn\\
&= -i(D-2)g^2 C_F \int {d^dk\over (2\pi)^d} \frac{p^0 - k^0}{k^2 (p-k)^2} \left(e^{ik\cdot b} - e^{ip\cdot b}\right)\,.\nn
\end{align}

Using Ward-identities, the Coulomb gauge part can be simplified as
\begin{align}
&\tilde{q}_2(b,p) \nn\\
&= -ig^2 C_F \int {d^dk\over (2\pi)^d} n\cdot (p-k)\frac{\mbox{Tr}\left[( \gamma^0 \slashed k \slashed n +\slashed n\slashed k \gamma^0 )\slashed p \right]/4}{k^2 (p-k)^2 (\vec{p}-\vec{k})^2}\nn\\
&\qquad\times  \left(e^{ik\cdot b} - e^{ip\cdot b}\right)\nn\\
&=-ig^2 C_F p^0\int {d^dk\over (2\pi)^d} (p^0-k^0)\frac{2(k^0+k^z)}{k^2 (p-k)^2 (\vec{p}-\vec{k})^2}\nn\\
&\qquad\times  \left(e^{ik\cdot b} - e^{ip\cdot b}\right)\nn\\
&=-ig^2 C_F p^0\int {d^dk\over (2\pi)^d} \left[ -{1\over (p-k)^2 (\vec{p}-\vec{k})^2} - {1\over k^2 (\vec{p}-\vec{k})^2} \right.\nn\\
&\qquad\qquad - {2\over k^2(p-k)^2} + {2(\vec{p}-\vec{k})\cdot \vec{p}\over k^2(p-k)^2(\vec{p}-\vec{k})^2} \nn\\
&\qquad\qquad \left.+ {2(p^0-k^0)k^z\over k^2 (p-k)^2 (\vec{p}-\vec{k})^2}\right]\left(e^{ik\cdot b} - e^{ip\cdot b}\right)\,,
\end{align}
and the gauge field self interaction part can be simplified as
\begin{align}
&\tilde{q}_3(b,p) \\
&= -ig^2 C_F \int {d^dk\over (2\pi)^d} \frac{\mbox{Tr}\left[ \gamma^0 \slashed p \right]/4}{(p-k)^2 (\vec{p}-\vec{k})^2} \left(e^{ik\cdot b} - e^{ip\cdot b}\right)\nn\\
&= -ig^2 C_F p^0 \int {d^dk\over (2\pi)^d} \frac{1}{(p-k)^2 (\vec{p}-\vec{k})^2} \left(e^{ik\cdot b} - e^{ip\cdot b}\right)\,.\nn
\end{align}

Interestingly, $\tilde{q}_3(b,p) $ cancels the first term in the square bracket of $\tilde{q}_2(b,p) $ for massless quarks ($p^0=p^z$). Adding all contributions together, we have
\begin{align}
&\tilde{q}_1(b,p) +\tilde{q}_2(b,p) + \tilde{q}_3(b,p) \\
&=\!-ig^2 C_F \!\int {d^dk\over (2\pi)^d} \left[  {(D-2)(p^0-k^0)-2p^0\over k^2(p-k)^2} \!-\! {p^0\over k^2 (\vec{p}-\vec{k})^2}  \right.\nn\\
&\qquad \left. + {2p^0(p_z^2 - k^0k^z)\over k^2(p-k)^2(\vec{p}-\vec{k})^2} \right]\left(e^{ik\cdot b} - e^{ip\cdot b}\right)\,.\nn
\end{align}

Each integral is computed as follows:
\begin{align}
    I_1&= -ig^2 C_F \int{d(b^zp^z)\over 2\pi}e^{-ix  p\cdot b}  \int {d^dk\over (2\pi)^d} \nn\\
    &\quad \times {(D-2)(p^0-k^0)-2p^0\over k^2(p-k)^2}\left(e^{ik\cdot b} - e^{ip\cdot b}\right)\nn\\
    &= -ig^2 C_F p^z\int {d^dk\over (2\pi)^d}{(D-2)(p^0-k^0)-2p^0\over k^2(p-k)^2} \nn\\
    &\quad \times e^{ik_\perp\cdot b_\perp}\left(\delta(k^z-xp^z) -\delta(p^z-xp^z)\right)\nn\\
    &-ig^2 C_F p^z\delta(1-x)\int {d^dk\over (2\pi)^d}{(D-2)(p^0-k^0)-2p^0\over k^2(p-k)^2}\nn\\
    &\qquad \times  \left(e^{ik_\perp \cdot b_\perp} -1\right)\nn\\
    &= {\alpha_sC_F\over 2\pi}\left[x\left({1\over\eps_{\rm ir}}+ L_b\right) + (1-x)\right]_+\theta(x)\theta(1-x) \nn\\
    &+ \delta(1-x){\alpha_sC_F\over 4\pi} \left( {1\over\eps_{\rm uv}} + L_b + 1 \right) + \ldots\,,
\end{align}
where $\ldots$ represents power or exponentially suppressed terms, and
\begin{align}
	L_b &= \ln{\mu^2 b_\perp^2 \over b_0^2}\,,\qquad b_0=2e^{-\gamma_E}\,.
\end{align}

\begin{align}
    I_2&= -ig^2 C_F \int{d(b^zp^z)\over 2\pi}e^{-ix z\cdot p} \nn\\
    &\qquad\times \int {d^dk\over (2\pi)^d} {-p^0\over k^2 (\vec{p}-\vec{k})^2} \left(e^{ik\cdot b} - e^{ip\cdot b}\right)\nn\\
    &= ig^2 C_F p^z\int {d^dk\over (2\pi)^d}  {p^0\over k^2 (\vec{p}-\vec{k})^2} e^{ik_\perp\cdot b_\perp} \nn\\
    &\qquad \times \left(\delta(k^z-xp^z) -\delta(p^z-xp^z)\right)\nn\\
    &\quad  +ig^2 C_F p^z\delta(1-x)\int {d^dk\over (2\pi)^d} {p^0\over k^2 (\vec{p}-\vec{k})^2} \left(e^{ik_\perp \cdot b_\perp} -1\right)\nn\\
    &=\ldots + \delta(1-x) {\alpha_s C_F\over 2\pi}\left(- {1\over\epsilon_{\rm uv}} - L_p -4 \right)\,,
\end{align}
where
\begin{align}
	L_p &= \ln{\mu^2\over 4p_z^2}\,.
\end{align}

\begin{align}
    I_3&= -ig^2 C_F \int{d(b^zp^z)\over 2\pi}e^{-ix z\cdot p} \nn\\
    &\qquad \times \int {d^dk\over (2\pi)^d} {2p^0(p_z^2 - k^0k^z)\over k^2(p-k)^2(\vec{p}-\vec{k})^2} \left(e^{ik\cdot b} - e^{ip\cdot b}\right)\nn\\
    &=- ig^2 C_F p^z\int {d^dk\over (2\pi)^d}  {2p^0(p_z^2 - k^0k^z)\over k^2(p-k)^2(\vec{p}-\vec{k})^2} \nn\\
    &\qquad \times e^{ik_\perp\cdot b_\perp}\left(\delta(k^z-xp^z) -\delta(p^z-xp^z)\right)\nn\\
    &\quad -ig^2 C_F p^z\delta(1-x) \int {d^dk\over (2\pi)^d} {2p^0(p_z^2 - k^0k^z)\over k^2(p- k)^2(\vec{p}- \vec{k})^2}\nn\\
    &\qquad \times  \left(e^{ik_\perp \cdot b_\perp} - 1\right)\nn\\
    &= {\alpha_sC_F\over 2\pi} \left[-{1+x\over 1-x}\left({1\over\eps_{\rm ir}}+ L_b\right)\right]_+\theta(x)\theta(1-x) + \ldots\nn\\
    &\quad + \delta(1-x){\alpha_sC_F\over 2\pi} \left[- {1\over 2} \ln^2 {4(p^zb_T)^2\over b_0^2} + 2 \ln {4(p^zb_T)^2\over b_0^2} \right.\nn\\
    &\qquad \qquad\qquad\qquad\qquad \left. - 8 + {\pi^2\over2}\right]\,.
\end{align}

Therefore, the one-loop quasi-TMD is
\begin{align}
&\tilde q^{(1)}(x, b_\perp,\mu, p^z,\eps) \\
&= I_1 + I_2 +I_3\nn\\
    &=  {\alpha_sC_F\over 2\pi} \left[-{1+x^2\over 1-x }\left({1\over\eps_{\rm ir}}+ L_b\right) + (1-x) \right]_+\theta(x)\theta(1-x) \nn\\
    &\quad + \delta(1-x){\alpha_sC_F\over 2\pi}\left[-{1\over2}{1\over \eps_{\rm uv}} + {1\over2}L_b - L_p- {1\over 2} \ln^2 {4(p^zb_T)^2\over b_0^2} \right.\nn\\
    &\qquad\qquad \left. + 2 \ln {4(p^zb_T)^2\over b_0^2} - {23\over2} + {\pi^2\over2}\right]\nn\,.
\end{align}

\subsection{Quasi soft function}

We use Collins' scheme to define the standard soft function, which uses a large rapidity $y_A$ as regulator. The Collins's scheme is characterized by two off-the-light-cone Wilson lines along the directions
\begin{align}
	n_A^\mu &= (n_A^+, n_A^-, n_A^\perp) = (1, - e^{-2y_A}, 0_\perp)\,,\qquad y_A \gg 1\,, \nn\\
	n_A^\mu &= (n_B^+, n_B^-, n_A^\perp) = (- e^{2y_B},1, 0_\perp)\,, \qquad y_B \ll -1\,.
\end{align}

According to the previous subsection, the quasi soft factor is given by
\begin{align}\label{eq:quasisoft}
	\tilde S_C(b_\perp, \mu, y_n) &=\lim_{y_A\to\infty} {S_C^0(b_\perp, \mu, y_A) \over S(b_\perp,\mu, y_A-y_n)}\,.
\end{align}

The one-loop Collins' soft function is~\cite{Ebert:2019okf}
\begin{align}\label{eq:csoft}
	&\lim_{y_A\to\infty}{S(b_\perp,\eps, y_A-y_n)\over S_0(b_\perp,\eps, y_A-y_B)} \\
	&= \lim_{y_A\to\infty, y_B\to-\infty}\sqrt{S_0(b_T, \eps, y_A-y_n) \over S_0(b_T, \eps, y_A-y_B) S_0(b_T, \eps, y_n-y_B)}\,,\nn
\end{align}
where each soft function $S_0$ uses an off-the-light-cone regulator, for example~\cite{Ebert:2019okf},
\begin{align}
	S_0(b_\perp,\mu, y_A-y_B) &=1+{\alpha_sC_F\over 2\pi}\left({1\over \eps_{\rm uv}} + L_b  \right)\nn\\
	&\qquad \qquad \times [2- 2(y_A-y_B)]\,.
\end{align}

The soft factor $S_0(b_\perp,\eps, y_A-y_B)$ appearing in \eq{csoft} is the zero-bin contribution to be subtracted from the bare Collins beam function which involves a staple-shaped Wilson line along the $n_B$ direction. Therefore, the overall soft function
\begin{align}
	&S(b_\perp,\eps, y_A-y_n) \nn\\
	&=\lim_{y_B\to-\infty}\sqrt{S_0(b_T, \eps, y_A-y_n)S_0(b_T, \eps, y_A-y_B) \over  S_0(b_T, \eps, y_n-y_B)}\nn\\
	& = \sqrt{S_0(b_T, \eps, (2y_A-2y_n)}\nn\\
 &=1+{\alpha_sC_F\over 2\pi}\left({1\over \eps_{\rm uv}} + L_b  \right) [1- 2(y_A-y_n)]\,.
\end{align}
The introduction of three Wilson lines in the soft function definition is not arbitrary. First of all, the zero bin cancels the rapidity divergences in the bare Collins beam function which depend on $y_B$ with $y_B\ll -1$, as well as the pinched-pole singularities in the self-energy of the Wilson lines with rapidity $y_B$. However, the zero-bin also introduces additional rapidity divergences and pinched-pole singularity from the Wilson lines of rapidity $y_A$, both of which need to be cancelled by the overall soft function. Therefore, the overall soft function must appear in the form of $\sqrt{S_0(b_T, \eps, (2y_A-2y_n)}$ instead of $S_0(b_T, \eps, y_A-y_n)$, as the latter would introduce extra pinched-pole singularities.

The Feynman rule for the one-loop zero-bin Coulomb-gauge quasi-soft function $S_C^0$ is~\cite{Ebert:2019okf}
\begin{align}
	&\delta S_C^0(b_\perp, \mu, y_A)\\
	 &= \int {d^d k\over (2\pi)^d} \mbox{PV}{g\tau_a n_A^\mu \over n_A\cdot k + i0} (-2\pi)\delta(k^2)\nn\\
	 &\qquad \times\pi_{\mu\nu}(k) \mbox{PV}{g\tau^a n_A^\nu \over n_A\cdot k - i0} \left(e^{ik_\perp\cdot b_\perp}-1\right)\nn\\
 &=-g^2C_F \int {d^d k\over (2\pi)^d} (2\pi) \delta (k^2)\left(e^{ik_\perp\cdot b_\perp}-1\right) \nn\\
 &\qquad \times \mbox{PV}{n_A^2 -  2n_A\cdot k{(n_A\cdot n)(n\cdot k)\over \vec{k}^2} + {(n_A\cdot k)^2 \over \vec{k}^2} \over (n_A\cdot k +i0) (n_A\cdot k-i0) }\nn\\
 &=-g^2C_F\! \int {d^d k\over (2\pi)^d} (2\pi) \delta (k^2)\left[\mbox{PV}{n_A^2\over (n_A\!\cdot\! k \!+\!i0) (n_A\!\cdot\! k \!-\! i0) } \right.\nn\\
 &\qquad \left.- 2{(n_A\cdot n)(n\cdot k)\over \vec{k}^2}\mbox{PV}{1\over n_A\cdot k} + {1\over \vec{k}^2}\right]\left(e^{ik_\perp\cdot b_\perp}-1\right)\,.\nn
\end{align}

Without the principal-value prescription, the first term inside the square brackets would include a pinched-pole singularity, which shall be cancelled in the ratio in \eq{quasisoft}. Therefore, in practice we evaluate it with the principal-value prescription~\cite{Ebert:2019okf}, which effectively ignores the pinched-pole singularity,
\begin{align}
   &\delta  S_{C,1}^0(b_\perp, \mu, y_A) \\
   &= -g^2C_F \int {d^d k\over (2\pi)^d} (2\pi) \delta (k^2) \mbox{PV}{n_A^2\over (n_A\cdot k +i0) (n_A\cdot k-i0) }\nn\\
   &\qquad \times \left(e^{ik_\perp\cdot b_\perp}-1\right)\nn\\
    &= {\alpha_sC_F\over 2\pi}\left({1\over \eps_{\rm uv}} + L_b\right)\,,\nn
\end{align}

\begin{align}
 & \delta   S_{C,2}^0(b_\perp, \mu, y_A) \\
 &=2g^2C_F (n_A\cdot n)\int {d^d k\over (2\pi)^d}(2\pi) \delta (k^2) {n\cdot k\over \vec{k}^2}\nn\\
 &\qquad \times \mbox{PV}{1\over n_A\cdot k}\left(e^{ik_\perp\cdot b_\perp}-1\right)\nn\\
    &=2g^2C_F \int {d^d k\over (2\pi)^d}(2\pi) \delta (k^2) {{k^0}\over\vec{k}^2}\mbox{PV}{1\over k^0 - \lambda k^z}\left(e^{ik_\perp\cdot b_\perp} \!-\!1\right)\nn\\
    &=g^2C_F \int {d^{d-1} \vec{k}\over (2\pi)^{d-1}} {\left(e^{ik_\perp\cdot b_\perp}-1\right)\over k_z^2 + k_\perp^2}\mbox{PV}{1\over \sqrt{k_z^2+k_\perp^2} - \lambda k^z}\nn\\
    &= g^2C_F {1\over \lambda}\ln{\lambda+1\over \lambda-1}  \int {d^{d-2} \vec{k}\over (2\pi)^{d-1}} {1\over k_\perp^2}\left(e^{ik_\perp\cdot b_\perp}-1\right)\nn\\
    &=- {\alpha_s C_F\over 2\pi} {1\over \lambda}\ln{\lambda+1\over \lambda-1} \left({1\over \eps_{\rm uv}} + L_b\right) \nn\\
    &\overset{y_A\to\infty}{=} {\alpha_s C_F\over 2\pi} \left({1\over \eps_{\rm uv}} + L_b\right)  (-2y_A)\,,\nn
\end{align}
where
\begin{align}
	\lambda &= {1+e^{-2y_A}\over1-e^{-2y_A} } >1 \,.
\end{align}

\begin{align}
   &\delta  S_{C,3}^0(b_\perp, \mu, y_B) \nn\\
   &=-g^2C_F \int {d^d k\over (2\pi)^d} (2\pi)\delta(k^2){1\over \vec{k}^2}\left(e^{ik_\perp\cdot b_\perp}-1\right)\nn\\
    &= {\alpha_s C_F\over 2\pi}\left({1\over \eps_{\rm uv}} + L_b \right)\,.
\end{align}

Therefore, the zero-bin contribution to the quasi-TMD is
\begin{align}
	S_C^0(b_\perp, \mu, y_B) &=1+ \delta S_{C,1}^0 + \delta S_{C,2}^0 + \delta S_{C,3}^0 \\
	&=1 + {\alpha_sC_F\over 2\pi}  \left({1\over \eps_{\rm uv}} + L_b \right) (2-2y_A)\,,\nn
\end{align}
and the full quasi soft factor
\begin{align}
	\tilde S_C(b_\perp, \mu, y_n) &=\lim_{y_A\to \infty} {S_C^0(b_\perp, \mu, y_A) \over S(b_\perp,\mu, y_n-y_A)} \\
	&=1+ {\alpha_sC_F\over 2\pi}\left({1\over \eps_{\rm uv}} + L_b \right)(1-2y_n)\,.\nn
\end{align}

We have also tested the $\eta$-regulator~\cite{Chiu:2011qc,Chiu:2012ir}, where the zero-bin contribution is trivially 1, and the one-loop soft function is~\cite{Ebert:2019okf}
\begin{align}
    S(b_\perp, \mu, \nu, \eta) &=1+ {\alpha_sC_F\over 2\pi}\left[{1\over \eps_{\rm uv}^2} + \left({1\over \eps_{\rm uv}} + L_b\right)\right.\\
    &\qquad \left.\times \left(-{2\over \eta} + \ln {\mu^2\over \nu^2}\right) - {1\over2}L_b^2 - {\pi^2\over 12}\right]\,.\nn
\end{align}
The zero-bin contribution to the quasi-TMD is
\begin{align}
	&S_C^0(b_\perp, \mu, \nu,\eta) \\
 &=1-g^2C_F \int {d^d k\over (2\pi)^d} (2\pi) \delta (k^2) \nn\\
 &\qquad \times \left[- {\sqrt{2} k^0\over \vec{k}^2}{1\over k^-}{(2k^0)^{-{\eta}} \over \nu^{-{\eta}}}  + {1\over \vec{k}^2}\right]\left(e^{ik_\perp\cdot b_\perp}-1\right)\nn\\
 &=1-g^2C_F\left[-\int {d^{d-1} k\over (2\pi)^{d-1}}\left({2\over \eta} {1\over k_\perp^2} - {\ln(k_\perp^2/\nu^2)\over k_\perp^2}\right) \right.\nn\\
 &\qquad \left. + \int {d^{d-1} k\over (2\pi)^{d-1}} {1\over 2(\vec{k}^2)^{3\over2}}\right]\left(e^{ik_\perp\cdot b_\perp}-1\right)\nn\\
 &=1+ {\alpha_s C_F\over 2\pi}\left[ {1\over \eps_{\rm uv}^2} + \left({1\over \eps_{\rm uv}} + L_b \right)\right.\nn\\
 &\qquad \left.\times \left(1 -{2\over\eta} + \ln{\mu^2\over \nu^2}\right) - {1\over 2}L_b^2  - {\pi^2\over 12} \right]\,.
\end{align}
As a result, the Coulomb-gauge quasi soft factor is
\begin{align}
	\tilde S_C(b_\perp, \mu, y_n=0) &={S_C^0(b_\perp, \mu, \nu, \eta) \over S(b_\perp,\mu, \nu, \eta)} \nn\\
	&=1+ {\alpha_sC_F\over 2\pi}\left({1\over \eps_{\rm uv}} + L_b \right)\,,
\end{align}
which corresponds to the case when $y_n=0$ and the Collins-Soper scale $\zeta= 4p_z^2$, and is consistent with that from the off-the-light-cone rapidity regulator in the Collins scheme.

\subsection{Perturbative matching}

The one-loop physical TMDPDF is~\cite{Ebert:2019okf}
\begin{align}
	&q^{(1)}(x, b_\perp,\mu, \eps)   \nn\\
	 &=  {\alpha_sC_F\over 2\pi} \left[-{1+x^2\over 1-x }\left({1\over\eps_{\rm ir}}+ L_b\right) + (1-x) \right]_+\theta(x)\theta(1-x) \nn\\
    &+ \delta(1-x){\alpha_sC_F\over 2\pi}\left[{1\over\eps_{\rm uv}^2}- {1\over2}L_b^2 + \left({1\over\eps_{\rm uv}}+L_b\right)\right.\nn\\
    &\qquad \left. \times \left({3\over2} + \ln{\mu^2\over \zeta}\right) + {1\over2} - {\pi^2\over12}\right]\,,
\end{align}
where $\zeta=4p_z^2 e^{-2y_n}$.

Therefore, in the $\MS$ scheme, we subtract all the $1/\eps_{\rm uv}$ poles, and find out that 
\begin{align}
	&\tilde q^{(1)} -\delta(1-x) \tilde S_C^{(1)} - q^{(1)} \nn\\
 &=\delta(1-x) {\alpha_sC_F\over 2\pi}\left[-{1\over 2}L_p-3L_p-12+{7\pi^2\over12}\right]\,,
\end{align}
which is free from the IR logarithm $L_b$, thus validating the factorization formula at one-loop.


\bibliography{Coulomb}

\end{document}